# Large-aperture Computational Single-sensor Microwave Imager Using 1-bit Programmable Coding Metasurface at Single Frequency

Dapeng Lao; Lianlin Li*, Senior Member; Jun Ding; Yun Bo Li; and Tie Jun Cui* , Fellow

*Abstract*—**The microwave imaging based on inverse scattering strategy holds important promising in the science, engineering, and military applications. Here we present a compressed-sensing (CS) inspired large- aperture computational single-sensor imager using 1-bit programmable coding metasurface for efficient microwave imaging, which is an instance of the coded aperture imaging system. However, unlike a conventional coded aperture imager where elements on random mask are manipulated in the pixel-wised manner, the controllable elements in the proposed scheme are encoded in a column-row-wised manner. As a consequence, this single-sensor imager has a reduced data-acquisition time with improved obtainable temporal and spatial resolutions. Besides, we demonstrate that the proposed computational single-shot imager has a theoretical guarantee on the successful recovery of a sparse or compressible object from its reduced measurements by solving a sparsity-regularized convex optimization problem, which is comparable to that by the conventional pixel-wise coded imaging system. The excellent performance of the proposed imager is validated by both numerical simulations and experiments for the high-resolution microwave imaging.**

## I. INTRODUCTION

**M**icrowave imaging is an important and powerful technique in science, engineering, and military.[1–3] In the context of high-frame-rate electromagnetic (EM) imaging, the obtainable temporal and spatial resolutions are mainly limited by the sustainable throughput of the imager's memory, exposure time, and illumination conditions. Over the past decade, the coded aperture imaging system in combination with the sparsity-regularized reconstruction algorithm has gained intensive attentions.[4–8] A coded aperture imaging architecture relies on the use of a sequence of random masks, through which the modulated information of the probed object was fully captured by a single fixed sensor. Then, the information of the probed object can be faithfully retrieved with reduced number of measurements by solving a tractable optimization problem. When the probed object allows for a low-dimensional representation in certain transformed domain, either pre-specified or trained, such as DCT, wavelet, etc., it is known that such a single-sensor imager benefits from a fundamental fact that the number of measurements could be drastically reduced compared to that required by conventional imaging techniques. Interestingly, the required measurements could be significantly less than the unknowns to be reconstructed, as claimed by the compressed sensing (CS) theory (e. g., Refs. [9–11]). In this area, a pioneering work is the well-known single-pixel camera invented in Rice University.[4] Basically, the working principle of the single-sensor imager for high-resolution imaging is described as follows. The wavefronts scattered from the probed object are firstly modulated by the random masks. Then, the modulated information is captured by a fixed single sensor. Finally, the information of the probed object is retrieved by a sparsity-regularized reconstruction algorithm. An essential issue of the single-sensor imager is the construction of multiple- mode modulators or masks, which is not well tackled and remains challenging in designing the controllable masks, especially in the microwave frequencies.

Metasurfaces have shown great promise in manipulating electromagnetic waves, including the microwave and millimeter waves in a flexible way, as evidenced by a number of interesting applications, such as ultrathin flat lens,[12–16] analogy signal processing,[17] high-resolution hologram,[18–20] and some other functional devices.[21–23] Therefore, metasurfaces have increasing abilities in designing the spatial modulator, and have become an important component of the cutting-edge imaging systems, especially for the single-sensor imager. More recently, the programmable coding metasurfaces [24,25] have been introduced to dynamically manipulate the EM waves in both microwave frequency and beyond.

Here, we propose a new method to realize the large-aperture single-sensor microwave imager by utilizing the 1-bit programmable coding metasurface composed of an array of voltage-controllable particles. Each metasurface particle illuminated by the incident wave could be in a state of two distinct responses: "1" for important radiation when the loaded voltage is at a high level, and "0" for almost negligible radiation. In this way, a sequence of different quasi-random radiation patterns are easily generated by managing the applied voltages of the coding metasurface, which could provide adequate modes in our imaging system. Compared with the systems of transforming frequency-dependent masks using dispersive and resonant metasurfaces,[26,27] the proposed programmable imaging system with varied modulators under single frequency could avoid the object dispersion. Although the proposed imager is an instance of the coded aperture imaging systems, it is different from the conventional CS-inspired imagers (e.g., the single-pixel camera[4] and a recent terahertz single-sensor imager[5]) where the elements of the random masks are manipulated in the pixel-wised manner. The controllable elements in this scheme are manipulated in the column-row-wised manner, which could greatly simplify the shutter control mechanism of the pixel-wised coded exposure. Besides, we show that the proposed large-aperture computational single-shot imager has a theoretical guarantee on successful recovery of a sparse or compressible object from its reduced measurements by solving a sparsity-regularized convex optimization problem, which is comparable to that by a conventional coded imaging system. Furthermore, we fabricate a sample of such a 1-bit coding metasurface and conduct the proof-of-concept imaging experiments in the microwave frequencies, validating the performance of the novel single-sensor imaging system.

## II. PROBLEM STATEMENT AND METHODOLOGY

As illustrated in **Figure 1,** the proposed single- sensor imager is composed of three parts, including a transmitter working at single frequency to launch the incident wave, a 1-bit programmable coding metasurface to generate the sequentially CS random masks for modulating spatial wavefronts emanating from the transmitter, and a single sensor fixed at some distance away from the metasurface to collect the wave fields scattered from the probed object. The metasurface consists of a two-dimensional array of 1-bit voltage-controllable coding particles; and each particle could be in a state of two distinct responses: "1" when loaded with a higher voltage for important secondary radiations, and "0" for almost negligible radiation. The coding elements in the proposed scheme are manipulated in the column-row-wised manner, rather than the pixel- wised manner in the conventional CS methods**.** Specifically, the coding metasurface with $N_x \times N_y$ particles is controlled by $N_x + N_y$ instead of $N_x \times N_y$ random binary sequences. In this way, the programmable coding metasurface is capable of producing quasi-random patterns in a very flexible and dynamic manner.

The metasurface with the $m$th coded pattern, illuminated by an $x$-polarized plane wave, as illustrated in **Figure 1**, produces approximately an $x$-polarized radiation field at the location $\boldsymbol{r}$:

$$E^{(m)}(\boldsymbol{r}) = \sum_{n_x=1}^{N_x} \sum_{n_y=1}^{N_y} \tilde{A}_{n_x,n_y}^{(m)} g(\boldsymbol{r}, \boldsymbol{r}_{n_x,n_y}) \quad (1)$$
$$m = 1,2, \dots, M$$

where $g\left(\boldsymbol{r}, \boldsymbol{r}_{n_x,n_y}\right) = \frac{exp\left(jk_0\left|r - r_{n_x,n_y}\right|\right)}{4\pi\left|r - r_{n_x,n_y}\right|}$ is the three-dimensional Green's function in free space, $k_0$ is the operation wavenumber, $M$ is the total number of the coded patterns, $\tilde{A}_{n_x,n_y}^{(m)} = A_{n_x,n_y}^{(m)} \exp\left(j\varphi_{n_x,n_y}^{(m)}\right)$ describes the induced $x$-polarized current with the amplitude $A_{n_x,n_y}^{(m)}$ and phase $\varphi_{n_x,n_y}^{(m)}$ on the $(n_x, n_y)th$ unit of the coding metasurface, and $\boldsymbol{r}_{n_x,n_y}$ is the coordinate of the $(n_x, n_y)th$ metasurface particle. In Equation (1), the double summation is performed over all pixels of the coding metasurface, where $n_x$ and $n_y$ denote the running indices of the particle of the coding metasurface along

the $x$- and $y$-directions, respectively. The probed object with contrast function $O(\boldsymbol{r})$, falling into the investigation domain V, is illuminated by the wave field of Eq. (1), giving rise to the following electrical field at $\boldsymbol{r}_d$:[1–3]

$$E^{(m)}(\boldsymbol{r}_d) =$$
$$\sum_{n_x=1}^{N_x} \sum_{n_y=1}^{N_y} \tilde{A}_{n_x,n_y}^{(m)} \int_V g(\boldsymbol{r}, \boldsymbol{r}_{n_x,n_y}) g(\boldsymbol{r}_d, \boldsymbol{r})O(\boldsymbol{r})d\boldsymbol{r} \quad (2)$$
$$m = 1,2, \dots, M$$

Note that the Born approximation has been implicitly used in Equation (2).[1] After introducing the following function

$$\tilde{O}_{n_x,n_y} = \int_V g(\boldsymbol{r}, \boldsymbol{r}_{n_x,n_y}) g(\boldsymbol{r}_d, \boldsymbol{r})O(\boldsymbol{r})d\boldsymbol{r} \quad (3)$$
$$n_x = 1,2, \dots, N_x, n_y = 1,2, \dots, N_y$$

Equation (2) becomes

$$E^{(m)}(\boldsymbol{r}_d) = \sum_{n_x=1}^{N_x} \sum_{n_y=1}^{N_y} \tilde{A}_{n_x,n_y}^{(m)} \tilde{O}_{n_x,n_y} \quad (4)$$
$$m = 1,2, \dots, M$$

By applying the far-field approximation, Eq. (3) can be expressed as

$$\tilde{O}_{n_x,n_y} = \left(\frac{1}{4\pi}\right)^2 \frac{\exp\left(jk_0 r_{n_x,n_y}\right)}{r_{n_x,n_y}} \frac{\exp\left(jk_0 r_d\right)}{r_d}$$
$$\times \int_V exp\left(-j(\boldsymbol{k}_{n_x,n_y} + \boldsymbol{k}_d) \cdot \boldsymbol{r}\right) O(\boldsymbol{r})d\boldsymbol{r} \quad (5)$$

where $\boldsymbol{k}_{n_x,n_y} = k_0 \frac{\boldsymbol{r}_{n_x,n_y}}{r_{n_x,n_y}}$ and $\boldsymbol{k}_d = k_0 \frac{\boldsymbol{r}_d}{r_d}$. Eq. (5) implies that $\tilde{O}_{n_x,n_y}$ corresponds to two-dimensional discrete Fourier transform of $O(\boldsymbol{r})$. Notice that the spatial bandwidth of $O(\boldsymbol{r})$ is limited by the maximum value of $|\boldsymbol{k}_{n_x,n_y} + \boldsymbol{k}_d|$, and is determined by the maximum size of the coding metasurface. In other words, it can be deduced that the achievable resolution on $O(\boldsymbol{r})$ is in the order of $O(\lambda R/D)$,[28] where $\lambda$ is the operating wavelength, $R$ is the observation distance, and $D$ is the maximum size of the coded aperture.

In the context of computational imaging, Equation (4) can be reformulated in the following compact form:

$$E^{(m)} = \langle \tilde{\boldsymbol{A}}^{(m)}, \tilde{\boldsymbol{O}} \rangle \quad (6)$$
$$m = 1,2, \dots, M$$

where the symbol $\langle \cdot \rangle$ denotes the matrix inner-product, the matrix $\tilde{\boldsymbol{A}}^{(m)}$ has the size of $N_x \times N_y$ with entries of $\tilde{A}_{n_x,n_y}^{(m)}$, and the matrix $\tilde{\boldsymbol{O}}$ with the size of $N_x \times N_y$ is populated by $\tilde{O}_{n_x,n_y}$. Equation (6) reveals that the resulting problem of the computational imaging consists of retrieving $N = N_x \times N_y$ unknowns $\{\tilde{O}_{n_x,n_y}\}$ from the $M$ measurements $\{E^{(m)}(\boldsymbol{r}_d)\}$. Typically, Equation (6) has no unique solution if $N > M$ due to its intrinsic ill-posedness. To overcome this difficulty, we pursue a sparsity-regularized solution to Equation (6) since we believe that the probed object $\tilde{\boldsymbol{O}}$ has a low-dimensional representation in certain transform domain denoted by $\boldsymbol{\Psi}$, i.e., $\boldsymbol{\Psi}(\tilde{\boldsymbol{O}})$ being sparse. Therefore, the solution to Equation (6) could be achieved by solving the following sparsity-regularized optimization problem:

$$min_{\boldsymbol{O}} \left[\frac{1}{2}\sum_{m=1}^M \left(E^{(m)} - \langle \tilde{\boldsymbol{A}}^{(m)}, \tilde{\boldsymbol{O}} \rangle\right)^2 + \gamma ||\boldsymbol{\Psi}(\tilde{\boldsymbol{O}})||_1\right] \quad (7)$$

where $\gamma$ is a balancing factor to trade off the data fidelity and sparsity prior.

It is desirable that the number of sequential measurements of a fast microwave imaging should be as less as possible, and that the shutter control mechanism of the coded exposure should be as simple as possible. Traditionally, the coding pixels of the random mask are controlled independently, which is usually inefficient in data acquisition, and thus limits the temporal and spatial resolutions. To break this bottleneck, we propose the column- row-wised coding metasurface, by which the temporal random modulation of a $N_x \times N_y$ pixel array could be controlled by $N_x + N_y$ rather than $N_x \times N_y$ random binary sequences as required by a pixel-wise coding exposure. Specifically, the $N_x$-length random binary sequences of $\{0,1\}^{N_x}$ (denoted by $\boldsymbol{r} = [r_1, r_2, \ldots, r_{N_x}]$) and the $N_y$-length random binary sequences of $\{0,1\}^{N_y}$ (denoted by $\boldsymbol{c} = [c_1, c, \ldots, c_{N_y}]$) are used to control the row and column pixels, respectively. The row and column binary control signals jointly produce the binary random coded exposure sequence at the pixel location of $(n_x, n_y)$. Thus, one random realization of a coded pattern reads $\overline{\boldsymbol{A}} = \boldsymbol{r}^T \boldsymbol{c}$ up to a constant multiplicative factor. In this design, only $N_x + N_y$ control signals are needed to achieve the randomly coded exposures with $N_x \times N_y$ pixels, which could drastically reduce the complexity and increase the filling factor.

We now demonstrate that our single-sensor imager based on the 1-bit column-row-wised coding metasurface has a theoretical guarantee on successful recovery of a sparse or compressible object from its reduced measurements by solving a sparsity-regularized convex optimization problem. The conclusion is summarized in ***Theorem 1***, and the proof of which is provided in the **Appendix**.

***Theorem 1***. With $M$, $N$ and $S$ defined as previously, a $S$-sparse $N$-length signal can be accurately retrieved with the probability not less than $2\exp(-C(logSlogN)^2)$, provided that the number of measurements with $M \geq C * SlogN/S$ is up to a polynomial logarithm factor, where $C$ is a constant depending only on $S$.

We present a set of Monte-Carlo simulation results to verify Theorem 1 of the sampling theory. In this set of simulations, we consider the signals of dimension $N = N_x \times N_y = 32 \times 32$ whose nonzero entries are drawn i.i.d. from a standard normal distribution and are located on a support set drawn uniformly at random. We vary the number of random masks from $M$= 8 to $M$= 1024 with the step of size 8. Among all 100 trials run for each pair of $M$ and $S$, the ones that yield relative errors no more than $10^{-3}$ are counted as successful reconstruction. For comparison, the random mask programmed in the pixel-wised manner is investigated as well. **Figures 2a** and **2b** show the phase transition diagrams of the estimation produced by Eq. (2) in terms of the number of masks $M$ and $S$ for the proposed column-row-wised and conventional pixel-wised coded masks, respectively. In these figures, the $x$- and $y$-axes correspond to the values of $M$ and $S$, respectively. We observe that the phase transition boundaries are almost linear in both models, agreeing with the relation between $M$ and $S$ suggested by Theorem 1. Although the phase transition boundary for the column-row-wised model is slightly lower (worse) than that for the pixel-wised model, the difference is not significant.

## III. RESULTS AND DISCUSSIONS

Full-wave numerical simulations are conducted to examine the performance of the proposed single- sensor imager. The simulated results are obtained by the commercial software, CST Microwave Studio 2012, along with MATLAB 7.3. In our simulations, the 1-bit programmable metasurface is composed of $20 \times 20$ voltage-controllable ELC particles, shown in **Figures 3a** and **3b**. Each ELC particle has the size of $6 \times 6mm^2$ printed on the top surface of the FR4-substrate with a dielectric constant of 4.3 and a thickness of 0.2mm, plotted in **Figure 3b**. The ELC particle is loaded with a pair of identical PIN biased diodes (SMP 1320-079LF), and the 1-bit states of "0" and "1" are controlled by the applied bias direct-current (DC) voltage: when the biased voltage is on a high level (3.3 V), this pair of diodes are 'ON'; when there is no biased voltage, the diodes are 'OFF'. The effective circuit models of the biased PIN diode at the ON and OFF states are illustrated in **Figure 3c**. To show this more clearly, we plot the transmission responses (S$_{21}$) of the ELC particle loaded with the voltage- controlled PIN diodes in **Figure 3d**, in which the effective circuit models of the biased PIN diode are incorporated into the CST Microwave Studio. This figure clearly shows that at the working frequency 8.3GHz, the ELC particle behaves as a '1' element when the diodes are ON, and as a '0' element when the diodes are OFF.

In these simulations, we use two sets of controlling signals for controlling the rows and columns of the coding metasurface separately. The controlling signals are realized by two sequences of 20-digit Rademacher random binary variables, so the total number of the control units is $N_x + N_y = 40$ instead of $N_x \times N_y = 400$. Three samples of simulated radiation patterns of such a 1-bit coding metasurface are illustrated in **Figure 4b**, and the corresponding coded patterns of the programmable metasurface are shown in **Figure 4a**, in which the yellow and green parts correspond to the ON and OFF particles, respectively. For the convenience of readers, the column- and row-controlling signals are also plotted on the left and top sides of each coded pattern. More specifically, the 20-digit controlling signals are:

(10111110001010100010; 00110001110110001010),

(01010101110001011100; 01111011010011101000),

and

(11010000000011000100; 01011100010100001110),

respectively, where the first and second arguments are for column and row controlling, respectively. In our simulations, the incident wave is normally illuminated on the coding metasurface, and the radiation field is monitored at the plane of z=10mm away from the metasurface.

After generating the controlling signals, we then examine the performance of the proposed single- sensor imager by considering the "P"- and "K"- shaped metallic samples. In numerical simulations, the object plane is at the plane z=0.18m (five operating wavelengths) away from the metasurface, and the investigation domain with the size of $100 \times 100$ mm$^2$ is uniformly divided into 40×40 sub-grids. Note that the imaging resolution is around 18×18 mm$^2$. After obtaining the simulated data from CST Microwave Studio, we realize the reconstruction using TV-regularized minimization technique[28] in MATLAB. The final reconstruction results for the "P"- and "K"-shaped objects are illustrated in **Figure 4c** and **4d**, respectively, with different measurement numbers of M=200, 400, and 600 from left to right, respectively. It can be observed that image quality is steadily enhanced by increasing the number of measurements (i.e., using more random patterns). A closer look into the results shows that the images for "K" are slightly worse than those for "P", because the image of "P" is sparser than that of "K". In addition, it is worth noticing that the linearized imaging model characterized by Eq. (2) only takes the so-called single-scattering effect into account, and neglects more realistic interactions between the object and the wave field, which is to some extent responsible for the non-perfect reconstructions. Nonetheless, the above results have verified the performance of the proposed single-sensor imager based on the 1-bit coding metasurface.

A set of proof-of-concept experiments have been conducted to verify the performance of the proposed single-sensor microwave imager. To accomplish this goal, we fabricate a sample of 1-bit programmable metasurface that is encoded in the column-row-wised manner, as shown in **Figure 5a**. In our imaging experiments, a vector network analyzer (VNA, Agilent E5071C) is used to acquire the response data by measuring the transmission coefficients (S21). More specifically, a pair of horn antennas are connected to two ports of the VNA through two 4m-long 50-Ω coaxial cables: one is used for launching the incident wave, and the other for collecting the response data emanated from the probed object, illustrated in **Figure 5b**. To suppress the measurement noise level, the average number and filter bandwidth in VNA are set to 10 and 10 kHz, respectively. The experiments are carried out in a microwave anechoic chamber with the size of $2 \times 2 \times 2 \text{m}^3$. Other relevant parameters are kept as the same as those adopted in the numerical simulations.

First, these experiments are conducted to show the ability of the proposed coding metasurface to generate the controllable radiation patterns. The biased voltages of the coding metasurface in both column and row distributions could be digitally controlled by toggling different triggers, which control the 'ON' and 'OFF' states of the biased PIN diodes, thereby the required '0' or '1' state of each metasurface particle could be realized. Therefore, different quasi-random radiation patterns could be achieved. Totally 1000 random radiation patterns are generated for the imaging purpose, and three of them are shown in **Figures 5d**, corresponding to the controlling signals in **Figure 4a** from left to right, respectively. These radiation patterns are obtained by scanning the electrical fields at 3 mm away from the coding metasurface with a 50Ω-coaxial SMA tip, followed by the near-field-to-far-field transformation. The measured radiation patterns resemble those predicted by the numerical simulations, despite a little mismatches arising from the measurement errors, parasite effects of the diodes, and other possible reasons, which implies that the proposed 1-bit column-row-wise coding metasurface can be utilized to generate incoherent random-like masks.

Second, two sets of imaging results are presented to demonstrate the performance of the proposed single-sensor imager. As done in the numerical simulations, the "P" and

"K"-shaped metallic objects are considered in the experiments, shown in **Figure 5c**. The reconstruction results for "P" ("K") are provided in **Figure 5e-g** (**Figure 5h-j**), considering different numbers of measurement M= 200, 400, and 600, respectively. Similar conclusions to the previous numerical simulations can be drawn immediately. The experimental imaging results clearly validate the feasibility of the proposed single-sensor imaging system based on the 1-bit column-row-wised programmable coding metasurface.

## IV. CONCLUSIONS

In conclusion, we presented a new single-sensor imager based on the 1-bit column-row-wised programmable coding metasurface for the high-rate- frame electromagnetic imaging. A sample of such a 1-bit coding metasurface was fabricated and the proof-of-concept imaging tests were conducted in microwave frequencies to validate the single-sensor system for the high-frame-rate imaging, which could pave a new avenue for capturing and tracking moving objects, with either very high or very low speeds. The proposed single-sensor imager features two advantages over the existing methods: 1) avoiding the object dispersion due to the single- frequency reconstruction without the frequency agility; 2) encoding the programmable metasurface in the column-row-wise manner instead of the pixel-wise manner to reduce drastically the data acquisition time, resulting in the improved temporal and spatial resolutions. We also demonstrated that the very simple 1-bit column-row-wised coding metasurface has a theoretical guarantee to ensure that the required measurement number is comparable to that for the conventional pixel-wise encoded masks while maintaining nearly the same imaging quality. The proposed method is not only applicable for far-filed imaging, but also for near-field imaging, which will be studied in our future work. The new imaging system may be extended to the terahertz frequencies.

## APPENDIX

Detailed here is the proof of ***Theorem 1*** in the main text. The proof here resembles that of Bahmani and Romberg,[28] Eftekhari *et al*.,[29] and Pournaghi and Wu,[30] which used a powerful theorem obtained in Ref. [31]. For readers' convenience, we repeat this theorem below to facilitate our proof.

**THEOREM 2**[31]. Let $\mathcal{A} \in \mathbb{C}^{M \times N}$ be a set of complex-valued matrices, and let $\boldsymbol{\varepsilon}$ be a random vector whose entries are i.i.d. zero-mean, unit variance random variables with the sub-Gaussian norm$\tau$.Set $d_F(\mathcal{A}) = \sup_{\mathcal{A} \in \mathcal{A}} ||\mathcal{A}||_F$,

$d_2(\mathcal{A}) = \sup_{\mathcal{A} \in \mathcal{A}} ||\mathcal{A}||_2$,

$E_1 = \gamma_2(\mathcal{A}, ||\cdot||_2)\{\gamma_2(\mathcal{A}, ||\cdot||_2) + d_F(\mathcal{A})\} + d_F(\mathcal{A})d_2(\mathcal{A})$,

$E_2 = d_2(\mathcal{A})\gamma_2(\mathcal{A}, ||\cdot||_2) + d_F(\mathcal{A})$,

and, $E_3 = d_2^2(\mathcal{A})$,

then, for $t > 0$, it holds that

$\log \mathbb{P}\left\{\sup_{\mathcal{A} \in \mathcal{A}} |||\mathcal{A}\boldsymbol{\varepsilon}||_2^2 - \mathbb{E}||\mathcal{A}\boldsymbol{\varepsilon}||_2^2 | \geq_\tau E_1 + t\right\} \leq_\tau min\left(\frac{t^2}{E_2^2}, \frac{t}{E_3}\right)$

To prove **Theorem 1** in the main text, we assume that $\left\|X\right\|_F = 1$ without loss of generality, and then reformulate Eq. (1) as:

$$\langle \mathbf{X}, \boldsymbol{\Phi}_i \rangle = Trace(\boldsymbol{X}^T \boldsymbol{c}_i \boldsymbol{r}_i^T) = \boldsymbol{r}_i^T \boldsymbol{X}^T \boldsymbol{c}_i \tag{S1}$$

which can be furthermore expressed as

$$\frac{1}{\sqrt{M}} \langle \boldsymbol{\Phi}_i, \mathbf{X} \rangle = d_i, \quad m = 1,2,\dots,M \tag{S2}$$

or

$$\begin{bmatrix} (\boldsymbol{Xr}_i)^T & & \\ & \ddots & \\ & & (\boldsymbol{Xr}_M)^T \end{bmatrix} \begin{bmatrix} \boldsymbol{c}_1 \\ \vdots \\ \boldsymbol{c}_M \end{bmatrix} = \begin{bmatrix} d_1 \\ \vdots \\ d_M \end{bmatrix} \tag{S3}$$

Following the guideline adopted in Refs.[1–3], it is readily shown that

$$d_F(\mathcal{A}) = \sup_{A \in \mathcal{A}} \|A\|_F = \frac{1}{M} \sup_{A \in \mathcal{A}} \sum_{i=1}^{M} \|\boldsymbol{Xr}_i\|_2^2 = 1 \tag{S4}$$

and

$$d_2(\mathcal{A}) \le \sqrt{S/M}. \tag{S5}$$

In Eq. (S4),
$\sum_{i=1}^{M} \|\boldsymbol{Xr}_i\|_2^2 = Trace \sum_{i=1}^{M} \boldsymbol{Xr}_i \boldsymbol{r}_i^T \boldsymbol{X}^T = MTrace(\boldsymbol{XX}^T) = M$,
where $\|X\|_F = 1$ is used in the last equation, and $\boldsymbol{r}$ is an i.i.d. zero-mean and unit variance random vector. In Equation (S4), we obtain the following conclusion:

$$\begin{aligned} d_2(\mathcal{A}) &= \sup_{A \in \mathcal{A}} \|A\|_2 = \sup_{A \in \mathcal{A}} \sqrt{\left\| \boldsymbol{AA}^T \right\|_2} \\ &= \frac{1}{\sqrt{M}} \sup_{A \in \mathcal{A}} max_i \|\boldsymbol{Xr}_i\|_2 \\ &\le \frac{\|\boldsymbol{r}_i\|_\infty}{\sqrt{M}} \|X\|_1 \\ &\le \sqrt{\frac{S}{M}} \|\boldsymbol{r}_i\|_\infty \|X\|_F \\ &= \sqrt{\frac{S}{M}} \|\boldsymbol{r}_i\|_\infty \end{aligned}$$

On the other hand, the upper bound of $\gamma_2(\mathcal{A}, \|\cdot\|_2)$ can be estimated as[29]

$$\gamma_2(\mathcal{A}, \|\cdot\|_2) \le C \sqrt{\frac{S}{M}} \log S \, log N \tag{S6}$$

Equation (S4), (S5) and (S6) lead to the conclusion that $P(\delta_S \ge \hat{\delta}) \le 2\exp(-C(logSlogN)^2)$ holds for $M \ge C\hat{\delta}^{-2}S(logSlogN)^2$, which implies that the measurement operator $\mathcal{A}$ satisfies the so-called S-RIP with the isometry constant $0 < \delta_S < 1$ with the probability not less than $2\exp(-C(logSlogN)^2)$, provided that the number of measurements $M \ge C\hat{\delta}^{-2}S(logSlogN)^2$, where C is a constant only depending on $\delta_S$. As a consequence, **Theorem 1** in the main text can be obtained immediately.

## REFERENCES

[1]   M. Pastorino, Microwave imaging, John Wiley & Sons, Press, 2010
[2]   Lianlin Li, Wenji Zhang, and Fang Li, Derivation and discussion of the SAR migration algorithm within inverse scattering problem: theoretical analysis, IEEE Trans. Geoscience and Remote Sensing, 48(1):415-422, 2010
[3]   M. Amin, Through-the-wall radar imaging, CRC Press, 2010
[4]   M. F. Duarte et al., Single-pixel imaging via compressive sampling, IEEE Signal Process. Mag 25, 83, 2008.

[5]   C. M. Watts, D. Shrekenhamer, J. Montoya, G. Lipworth, J. Hunt, T. Sleasman, S. Krishna, D. R. Smith and W. J. Padilla, Terahertz compressive imaging with metamaterial spatial light modulators, Nat. Photonics. 8, 605, 2014.
[6]   Y. Li, Lianlin Li, B. Xu et al., Transmission-type 2-bit programmable metasurface for single-sensor and single-frequency microwave imaging, Scientific Reports, 2015
[7]   Chan, W. L., Charan, K., Takhar, D., Kelly, K. F., Baraniuk, R. G. & Mittleman, D. M. A single-pixel terahertz imaging system based on compressed sensing. Appl. Phys. Lett. 93, 121105 (2008).
[8]   Shrekenhamer, D., Wats, C. M. & Padilla, W. J. Terahertz single pixel imaging with an optically controlled dynamic spatial light modulator. Optics Express 21, 12507-12518 (2013).
[9]   E. Candes, J. Romberg, and T.Tao. Robust uncertainty principles: Exact signal reconstruction from highly incomplete frequency information. IEEE Trans. Inform. Theory, 52(2):489-509, 2006
[10]  E. Candes, J. Romberg, and T.Tao. Stable signal recovery from incomplete and inaccurate measurements. Comm. Pure Appl. Math., 59(8):1207-1223, 2006
[11]  D. Donoho, Compressed sensing. IEEE Trans. Inform. Theory, 52(4):1289-1306, 2006
[12]  C. L. Holloway, M. A. Mohamed, E. F. Kuester and A. Dienstfrey, Reflection and transmission properties of a metafilm: With an application to a controllable surface composed of resonant particles, IEEE Trans. Antennas Propag. 47, 853, 2005.
[13]  N. Yu, et al, Light Propagation with Phase Discontinuities: Generalized Laws of Reflection and Refraction, Science 334, 333, 2011.
[14]  N. Yu and F. Capasso, Flat optics with designer metasurfaces, Nat. Mater. 13, 139 , 2014.
[15]  L. X. Liu, et al, Broadband Metasurfaces with Simultaneous Control of Phase and Amplitude, Adv. Mater. 26, 5031, 2014.
[16]  C. Pfeiffer and A. Grbic, Metamaterial Huygens' Surfaces: Tailoring Wave Fronts with Reflectionless Sheets, Phys. Rev. Lett. 110, 197401, 2013.
[17]  Silva, A., & Engheta, N. Performing mathematical operations with metamaterials. Science 343, 160-3, 2014.
[18]  X. Ni, A. V. Kildishev and V. M. Shalaev, Metasurface holograms for visible light, Nat. Commun. 4, 2807, 2013.
[19]  L. Huang, et al, Three-dimensional optical holography using a plasmonic metasurface, Nat. Commun. 4, 2808, 2013.
[20]  G. Zheng, H. Mühlenbernd, M. Kenney, G. Li, T. Zentgraf and S. Zhang, Metasurface holograms reaching 80% efficiency, Nat. Nanotech. 10, 308, 2015.
[21]  Y. B. Li, X. Wan, B. G. Cai, Q. Cheng and T. J. Cui. Frequency-controls of electromagnetic multi-beam radiations and beam scanning by metasurfaces. Sci. Rep. 4, 6921, 2014.
[22]  S. Jiang, et al, Controlling the Polarization State of Light with a Dispersion-Free Metastructure, Phys. Rev. X. 4, 041042, 2014.
[23]  C. D. Giovampaola and N. Engheta, Digital metamaterials, Nat. Mater. 13, 1115 (2014).
[24]  T. J. Cui, M. Q. Qi., X. Wan, et al., Coding metamaterials, digital metamaterials and programmable metamaterials, Light: Science & Applications, 3, e218, 2014
[25]  T. J. Cui, S. Liu, and Lianlin Li, Information entropy of coding metasurface, Light: Science & Applications, 3, e218, 2016
[26]  J. Hunt et al., Metamaterial apertures for computational imaging, Science 339, 310 2013.
[27]  G. Lipworth, A. Mrozack, J. Hunt, D. L. Marks, T. Driscoll, D. Brady and D. R. Smith, Metamaterial apertures for coherent computational imaging on the physical layer, J. Opt. Soc. Am. A 30, 1603,2013.
[28]  S. Bahmani, and J. Romberg, Compressive deconvolution in random mask imaging, IEEE Trans. Computational Imaging, 1(4):236-246, 2015
[29]  A. Eftekhari, H. L. Yap, C. J. Rozell, and M. B. Wakin, The restricted isometry property for random block diagonal matrices, Appl. Comput., Harmonic Anal., 38, 1, 1-31, 2015
[30]  R. Pournaghi and X. Wu, Coded acquisition of high frame rate video, IEEE Trans. Image Processing, 23(12):5670-5681, 2014
[31]  F. Krahmer, S. Mendelson, and H. Rauhut, Suprema of chaos processes and the restricted isometry property, Communications on Pure and Applied Mathematics, LXVII, 187701904, 2014
[32]  S. H. Chan, R. Khoshabeh, K. B. Gibson et al., An augmented lagrangian method for total variation video restoration, IEEE Trans. Image Processing, 20, 11, 3097-3111, 2011
[33]  W. C. Chew, Waves and fields in inhomogeneous media, John Wiley & Sons Press, 1999

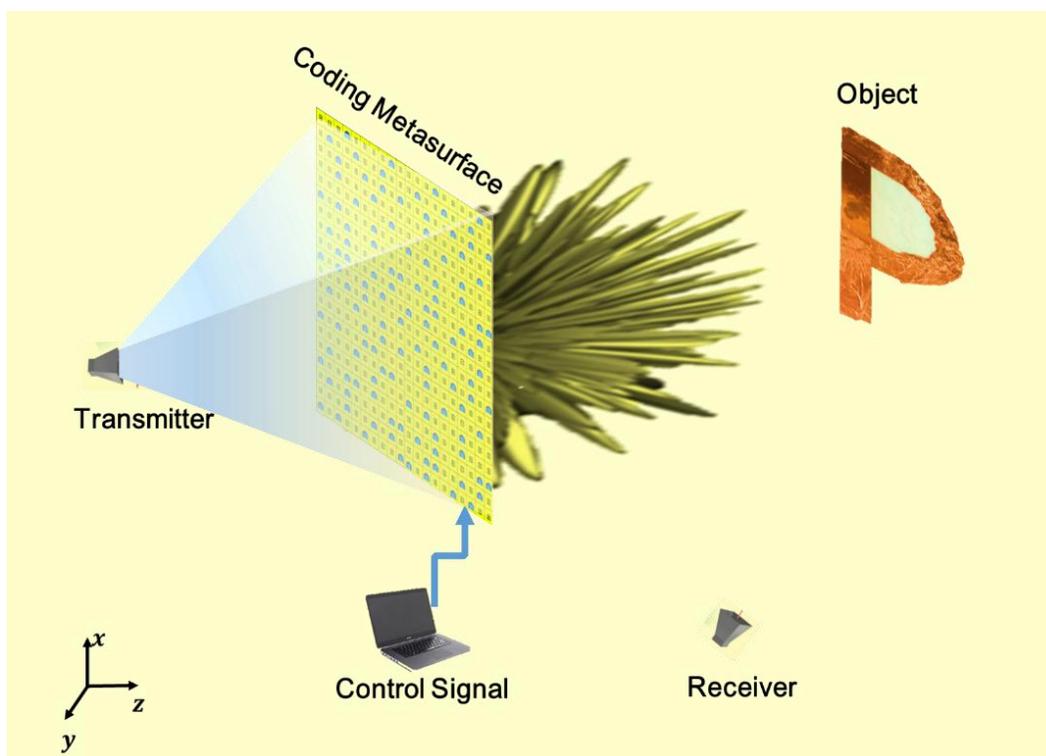

**Figure 1**. The schematic of the proposed single-sensor imager: this imager consists of a transmitter working with single frequency launches an illumination wave, a one-bit coding metasurface is responsible for generating sequentially random masks for modulating the spatial wavefront emerged from transmitter, and a single sensor is fixed at somewhere for collecting the wave field scattered from the probed object.

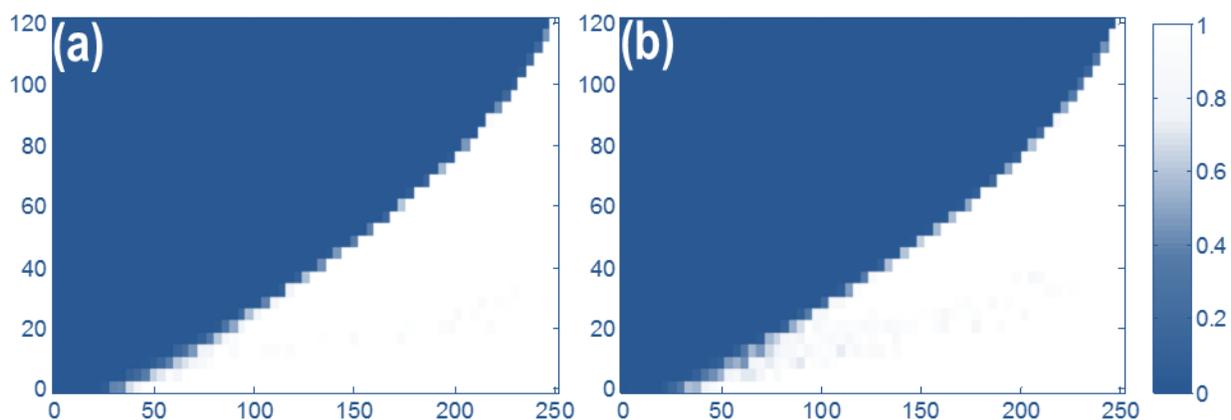

**Figure 2**: The phase transition diagram of l1-minimization in terms of M, the number of masks, and S, the l0-norm of the signals with dimension L = 1024. (**a**) The column-row-wised coded system. (**b**) The pixel-wise coded system. In these figures, the *x*-axis denotes M while the *y*-axis corresponds to S.

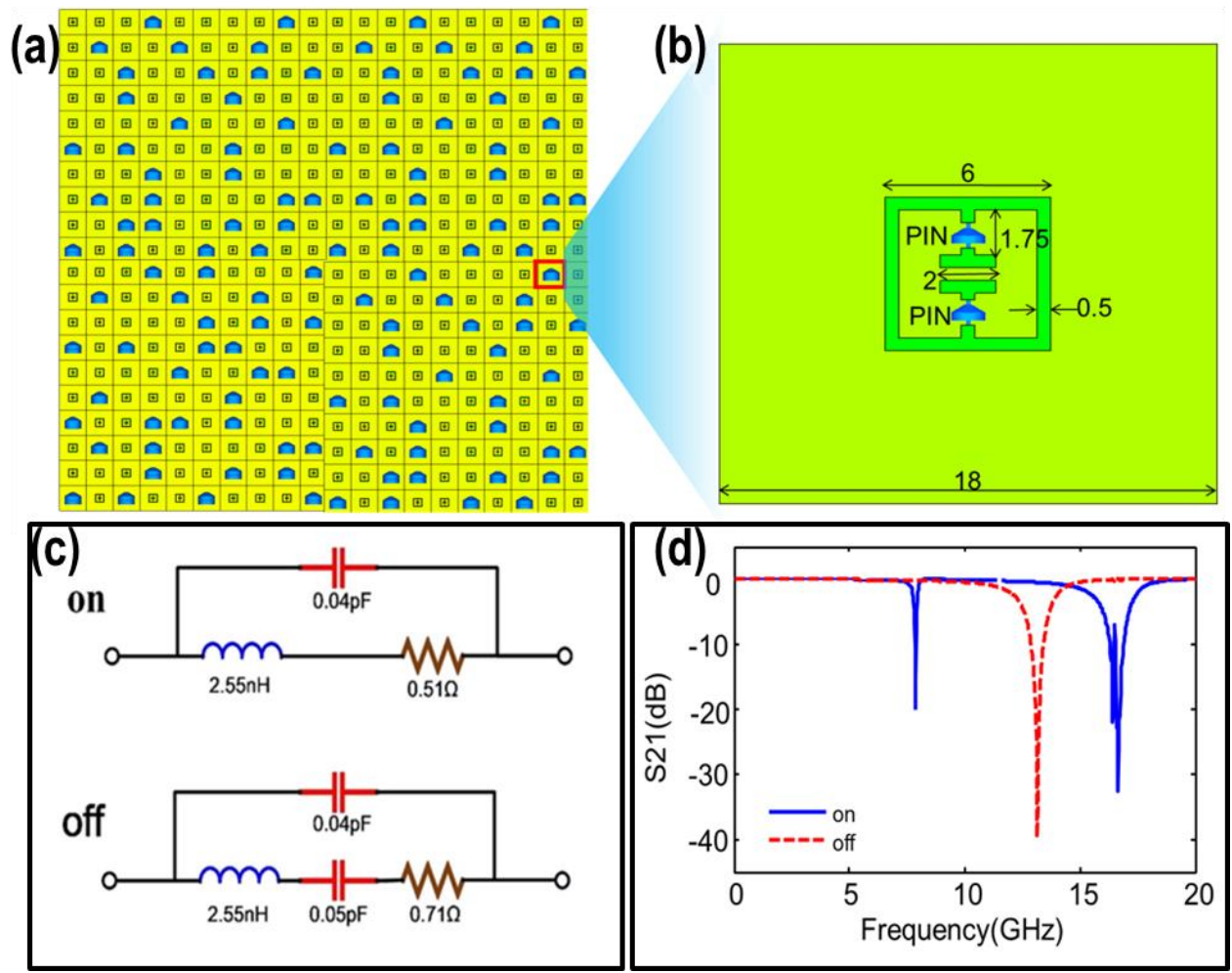

**Figure 3**. (**a**) The diagram of the 1-bit coding metasurface composed of 20×20 voltage-controlled ELC particles. For the visual purpose, the ELC particles coded with ON state are highlighted in blue, and others are not highlighted. (**b**) The configuration for the voltage-controlled ELC unit, which has a period of 18 mm, and a size of 6×6mm². The ELC unit is printed on a commercial printed circuit board FR4 with the relative permittivity of 4.3 and the thickness of 0.2 mm. (**c**) The effective circuit models of the biased diode at the ON and OFF states. (**d**) The S21 responses of the ELC particle loaded with PIN diode, implying that the ELC particle behaves as a '1' element when the diode is on, and as a '0' element when the diode is off at the working frequency 8.3GHZ.

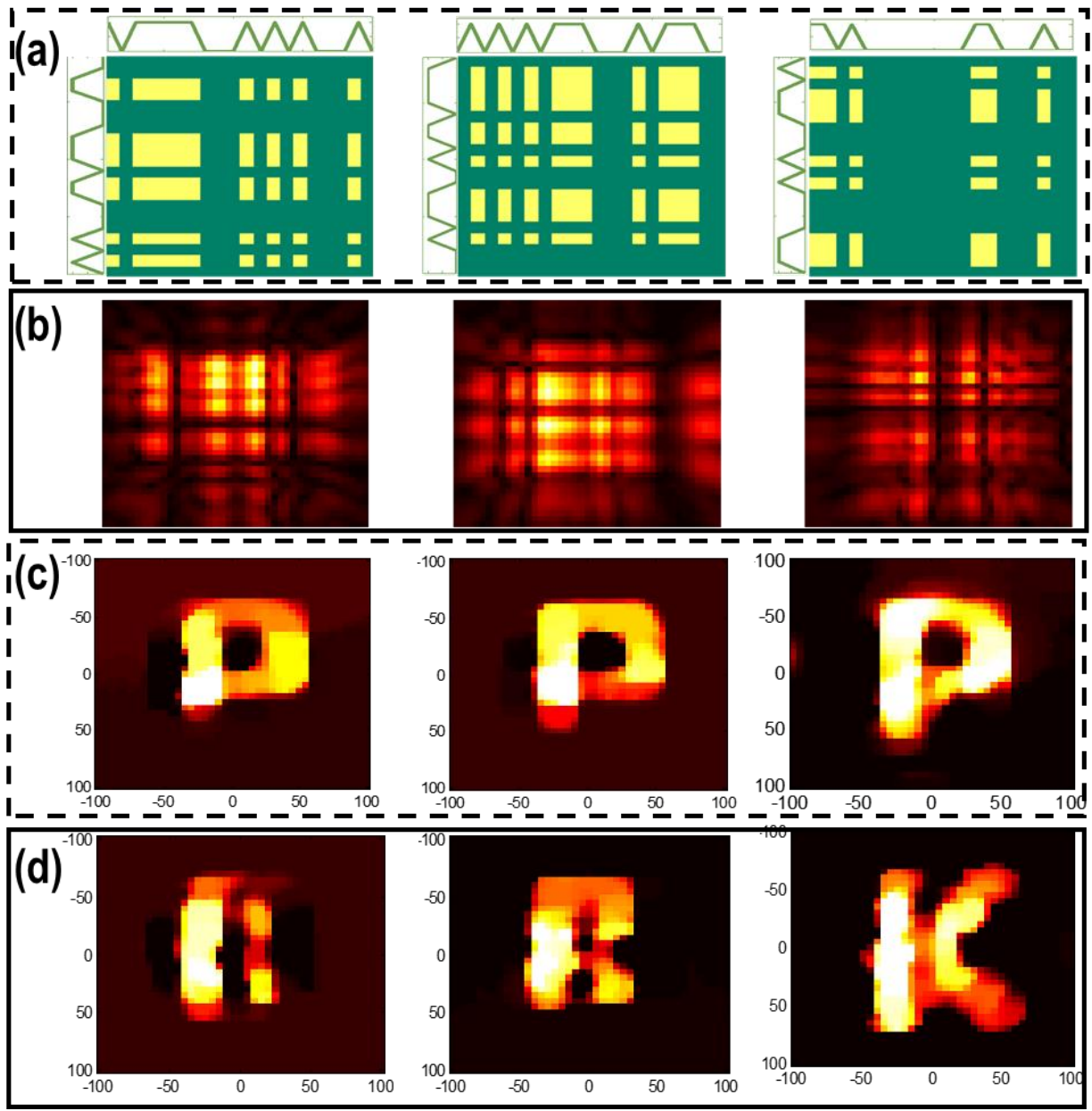

**Figure 4**. **(a)** Three coding patterns of the 1-bit column-row-wise encoded metasurface, in which the corresponding controlling signals for the column and row pixels are plotted in the left and top sides of each subfigure. Here, the yellow and green parts correspond to the "ON" and "OFF" elements, respectively. **(b)** Three radiation patterns of the metasurface corresponding to the coded patterns shown in (a). **(c)** The reconstruction results of the P-type metallic object for different measurement numbers M=200, 400, and 600, from the left to the right. **(d)** The reconstruction results of the K-type metallic object for different measurement numbers M=200, 400, and 600 from the left to right.

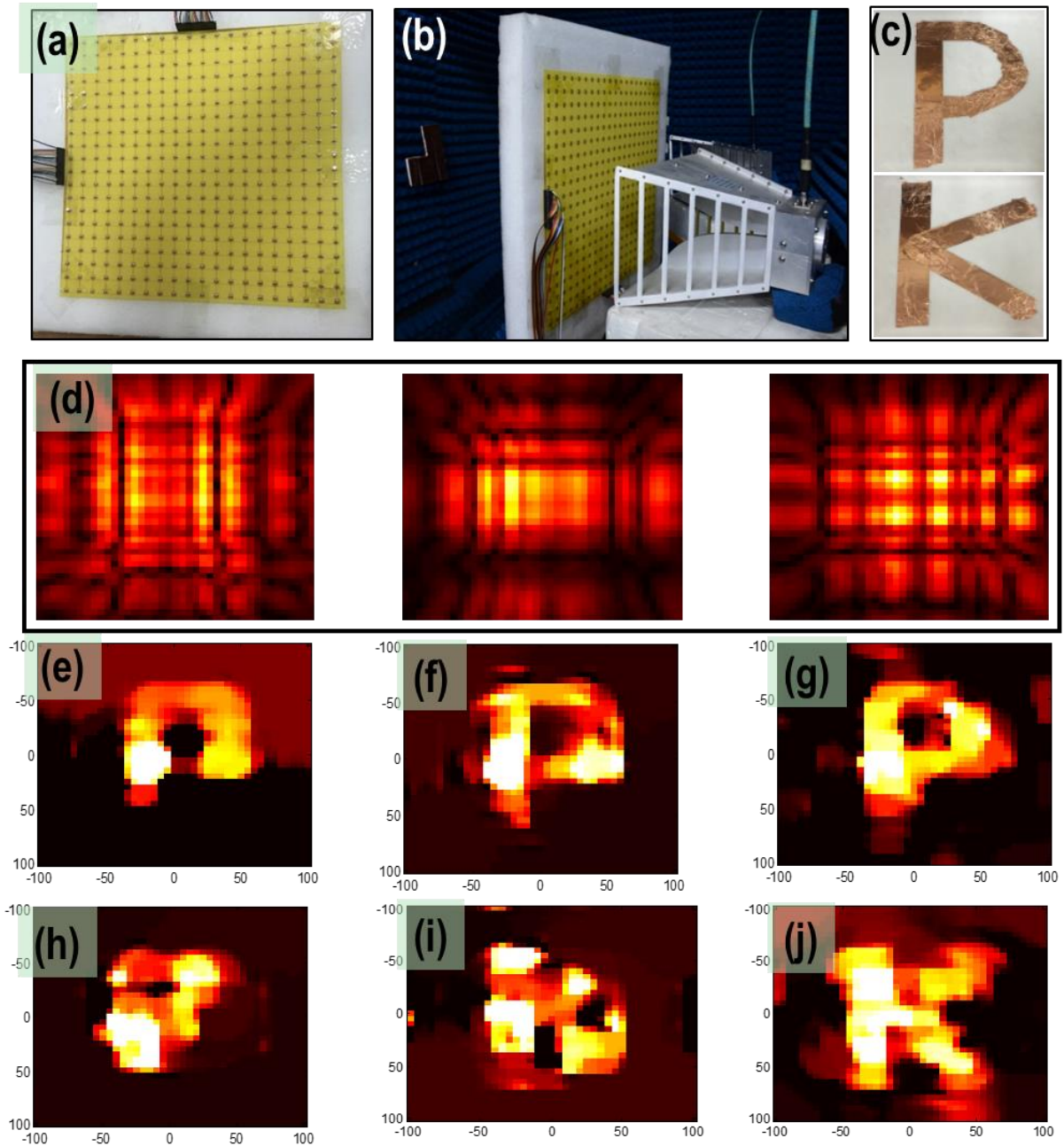

**Figure 5**. (**a**) The fabricated sample of the column-row-wise coding metasurface. (b) The experimental single-sensor imaging system based on the 1-bit programmable coding metasurface. (c) the "P"- and "K"-shaped metallic objects for imaging test. (**d**) Three samples of the radiation patterns of the metasurface encoded with the controlling signals used in **Figure 4a**. (**e-g**) The measured imaging results for the "P"-shaped metallic object with different measurement numbers M=200, 400, and 600, respectively. (**h-j**) The measured imaging results for the "K"-shaped metallic object with different measurement numbers M=200, 400, and 600, respectively.